\documentclass[12pt,preprint]{emulateapj}

\newcommand{\kms}{km\ s$^{-1}$}

\newcommand{\Msun}{$M_{\sun}$}

\newcommand{\Ha}{H$\alpha$\ }

\shorttitle{OVRO Observations of Spiral Arm Spurs in M51}
\shortauthors{Corder et al.} 

\begin{document}

\title{Detection of Dense Molecular Gas in Inter-Arm Spurs in M51}
 
\author{Stuartt Corder \altaffilmark{1}, Kartik Sheth
  \altaffilmark{1,2}, Nicholas Z. Scoville \altaffilmark{1}, Jin Koda
  \altaffilmark{1}, Stuart N. Vogel \altaffilmark{3}, Eve Ostriker
  \altaffilmark{3}}

\altaffiltext{1}{Department of Astronomy, California Institute of Technology, MS 105-24 Caltech, Pasadena, CA 91125}

\altaffiltext{2}{Spitzer Science Center \& Department of Astronomy, California Institute of Technology, MS 220-6 Caltech, Pasadena, CA 91125}

\altaffiltext{3}{Department of Astronomy, University of Maryland, College Park, MD 20742-2421}

\begin{abstract}

  Spiral arm spurs are prominent features that have been observed in
  extinction and 8$\mu$m emission in nearby galaxies.  In order to
  understand their molecular gas properties, we used the Owens Valley
  Radio Observatory to map the CO(J=1--0) emission in three spurs
  emanating from the inner northwestern spiral arm of M51.  We report
  CO detections from all three spurs.  The molecular gas mass and
  surface density are M$_{H2} \sim$3$\times$10$^6$ M$_{\sun}$ and
  $\Sigma_{H2} \sim$50 M$_{\sun}$ pc$^{-2}$.  Thus, relative to the
  spiral arms, the spurs are extremely weak features.  However, since
  the spurs are extended perpendicular to the spiral arms for
  $\sim$500 pc and contain adequate fuel for star formation, they may
  be the birthplace for observed inter-arm HII regions. This reduces
  the requirement for the significant time delay that would be
  otherwise needed if the inter-arm star formation was initiated in
  the spiral arms.  Larger maps of galaxies at similar depth are
  required to further understand the formation and evolution of these
  spurs and their role in star formation - such data should be
  forthcoming with the new CARMA and future ALMA telescopes and can be
  compared to several recent numerical simulations that have been
  examining the evolution of spiral arm spurs.

\end{abstract}

\section{Introduction}\label{intro}

One of the most striking features of the M51 mosaics from observations
with Hubble Space Telescope's (HST) Wide Field Planetary Camera and
Advanced Camera for Surveys (ACS) is the presence of regularly-spaced,
dark spurs extending perpendicularly outwards from the spiral arms
\citep{scoville01}.  With dust emission maps from the Spitzer Space
Telescope, the spurs are also seen most prominently in the IRAC
8$\mu$m maps of nearby galaxies (e.g., Figure 1 of
\citealt{brunner08}).  Additionally the IRAC images also reveal a
complex web of emission features which criss cross the spiral arms and
the inter-arm regions.  The 8$\mu$m data primarily trace polycyclic
aromatic
hydrocarbon (PAH) emission that is believed to originate from surfaces
of photo-dissociation regions (PDRs).  The exact nature of these
features is not well known but these features have been noted in
previous observational studies both in the Milky Way and in nearby
galaxies.  For instance, our own solar system is believed to lie in a
spur that is an offshoot of the Sagittarius spiral arm
\citep{weaver70}.  In a study of seven nearby spirals, including M51,
\citet{elmegreen80} measured an average pitch angle of $\sim$60
degrees and a width of $\sim$560 pc for spurs and concluded that they
were likely to be long-lived features.  These results were confirmed
in the higher-resolution, larger sample of \citet{lavigne06}.
\citet{elmegreen79} also noted that spurs are often accompanied by a
parallel string of HII regions and may trigger star formation.  In
barred spirals \citet{sheth00, sheth02} found a correlation between
spurs, which are normally seen upstream of bar dust lanes and HII
regions on the leading (downstream) side of the dust lanes.  They have
argued that spurs are conducive to star formation because of their
increased gas surface density relative to the surroundings and lower
shear relative to the bar dust lane.  Note however that the bar dust
lane spurs are usually on the trailing side of the dust lane, whereas
the spiral arm spurs are on the leading side, which suggests that they
may be a different phenomenon and have a different origin than the
spiral arm spurs.  In the most recent survey of nearby spirals,
\citet{lavigne08} find that spurs and feathers are common in disks
with well-defined dust lanes and the spur separation increases with
galactocentric radius.  They find spurs in 20\% of all nearby disks.
Recent numerical models have shown that spurs form downstream from
spiral arms under a variety of conditions
\citep{kim02,kim06,wada04,dobbs06, shetty06, wada08}.


In this paper we present observations of CO (J=1--0) emission line in
the inner, northwestern spiral arm and spurs and measure the masses
and average gas surface densities of the spurs.  The millimeter data
are the deepest observations of a field in M51 with the Owens Valley
Radio Observatory's (OVRO) millimeter array.  These data and analysis
provide the groundwork for the expected large surveys with CARMA and
ALMA which allow one to study galactic substructure over entire galaxy
disks in nearby galaxies.

\section{Observations and Data Reductions}\label{obs_red}
Single-pointing, OVRO $^{12}$CO (J=1--0, $\lambda$ 2.6~mm)
observations were carried out between January 2001 to April 2003, in
the compact, low and high resolution configurations.  These
configurations resulted in projected baselines ranging from 4 to 92
k$\lambda$.  The digital spectral correlator was configured for a
resolution of 2.6 \kms with a total bandwidth of 332 \kms.  The quasar
1156+295 was observed every 15--20 minutes for phase and amplitude
gain calibration.  The total on-source integration time was 54.5 hrs.
We reduced the data using the MMA software package \citep{scoville93},
and imaged the data using MIRIAD \citep{sault95}.  The synthesized
beam using robust weighting \citep{briggs95} is 2$\farcs$76$\times$
2$\farcs$10.  The FWHM of the primary beam is 1$\farcm$. The spectral
data was binned into 5 \kms channels for our analysis.  The rms noise
in a 5 \kms channel was 13 mJy/beam. Table \ref{m51props} lists
parameters for M51 parameters used in this paper and the pointing
center of the single OVRO field.

\begin{deluxetable}{crr}
\tabletypesize{\scriptsize} 
\tablecaption{M51 Properties \label{m51props}} 
\tablewidth{0pt} 
\tablehead{ \colhead{Parameter} &
\colhead{Value}
}
\startdata 
Center Position (2000.0) & $\alpha 13^h29^m50^s.38$, $\delta 47^o12'02''.28$\\
Distance$^{a}$ & 8.4 Mpc \\
Position Angle & $170^o$ \\
Inclination & $20^o$ \\
Linear Resolution & $112 \times 85$ pc  \\
Position Angle & $-57.9^o$ \\
Mass Conversion & $M = 1.5\times 10^4 S\Delta vD^2M_\sun$ \\
\enddata
\tablenotetext{a}{The value here is taken from \citet{Regan01} but
other values are reported in the literature which indicate an
uncertainty of 25\%.}
\end{deluxetable}

\section{Results}\label{results}

\begin{figure*}
\includegraphics*[width=15.0 cm, angle=-90]{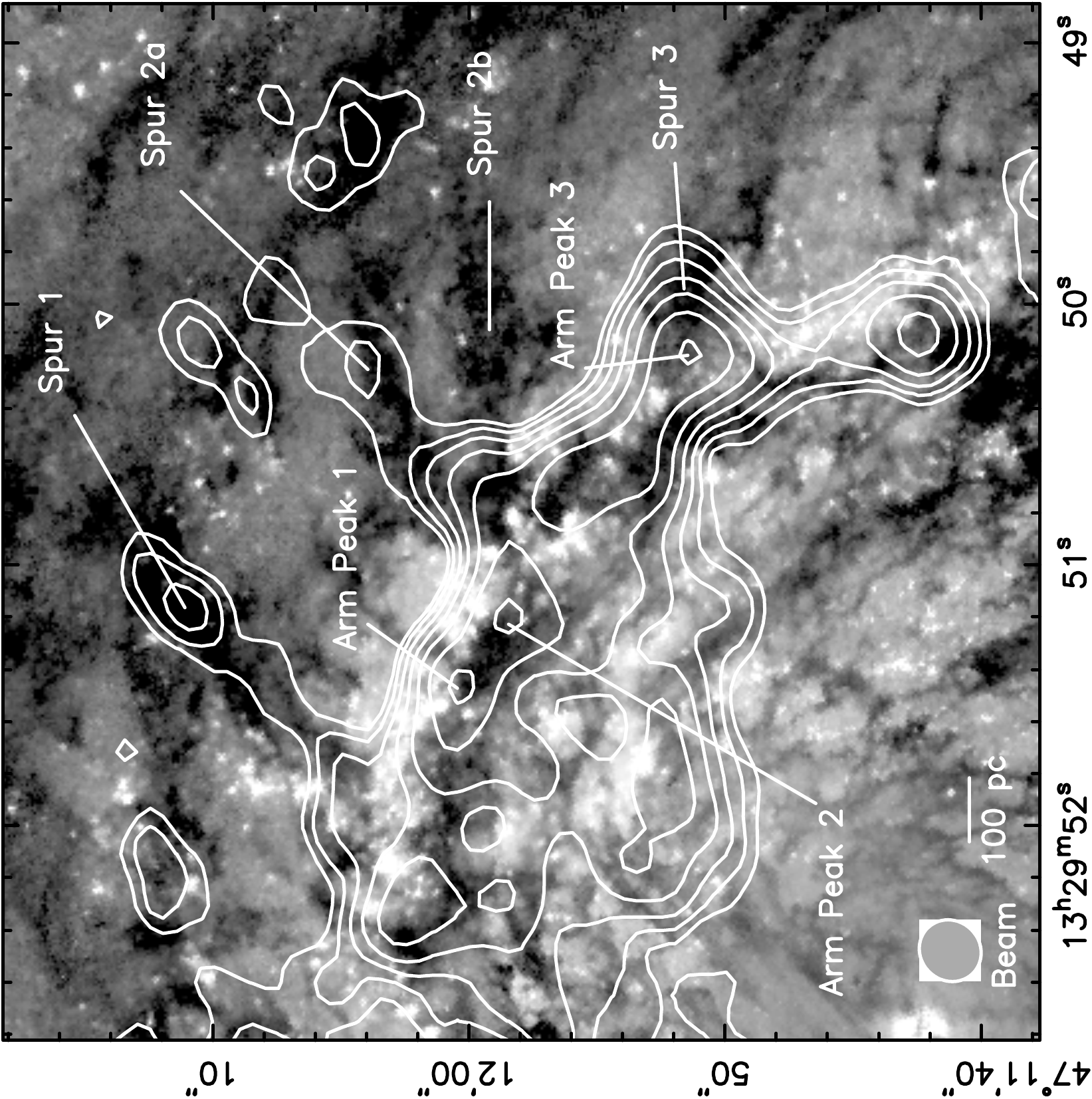}
\caption{V-band HST image of \citet{scoville01} with the velocity
  integrated CO map from OVRO overlaid in contours.  The contours are
  logarithmically spaced at 1 Jy bm$^{-1}$ km s$^{-1} \times $
  1.5$^{n}$, where n = 0,1,2,3,...6. The three spurs in the OVRO field
  are labeled as spurs 1, 2a and 3.  The section labelled spur 2b is a
  spur-like region but does not have associated CO emission.  
  The three peaks used to determine density contrasts of spiral arm peaks
  to the average spiral arm are labelled as arm peaks 1, 2, and 3.
  \label{ovroV}}
\end{figure*}

\begin{figure*}
\includegraphics*[width=17.0 cm]{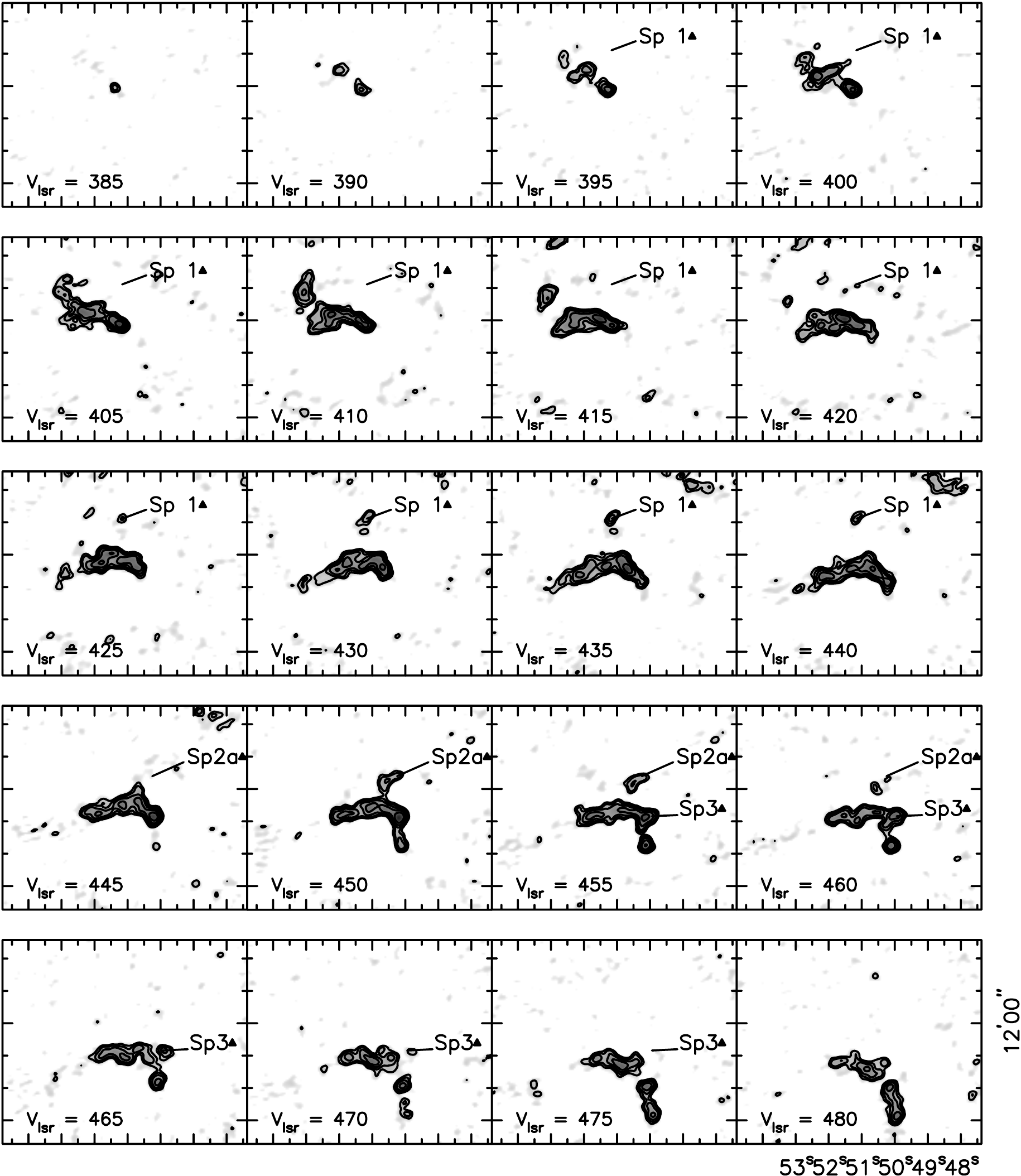}
\caption{Channels maps for spurs from the OVRO data.  The channels are
5 km s$^{-1}$ wide.  Spurs 1, 2a and 3 are labelled as Sp1, Sp2a and
Sp3 respectively.  Grey scale emission starts at $2\sigma$. The
contours are at 1.5$^n$ $\times$ 15 mJy beam$^{-1}$, with n=3,4,5,....
The emission peak at the tip of spur 1 is at 11$\sigma$ while the base
of this spur is seen at $\sim$ 5$\sigma$.  Spur 2a is detected at 9.5
$\sigma$ and spur 3 has emission between 4--10$\sigma$, although
separating spur from arm emission is difficult here.  The peak of each
spur is a about 2-3.5$\times$ lower than the peak emission in the arm
in the same channel.
  \label{chanmaps}}
\end{figure*}

The velocity integrated molecular gas emission is shown overlaid on a
V-band HST image in Figure \ref{ovroV} for the field targeted by our
OVRO observations.  The optical image shows four spurs labeled Spur 1,
2a, 2b and 3.  Molecular gas emission is associated with three (Spur
1, 2a and 3) of these four spurs.  Two of the three spurs have very
similar morphology, with an dense tip and an extended body which
connects back to the base at the spur/arm interface. Additional
features in the interarm region are not labeled.  The CO emission in
the spiral arm is extremely bright with a steep gradient on the
upstream and downstream edges.  Peak emission in the arm is an order
of magnitude brighter than in the spurs.  The brightest spiral arm
emission is in a ridge on the downstream side of the arm with distinct
peaks of gas surface density that are labeled as Arm Peak 1,2, and 3.
The channel maps for the CO emission are shown in Figure
\ref{chanmaps}, where each of the spurs can be uniquely identified in
a distinct set of channels (Spur 1 from 380--440 \kms\, Spur 2a from
430--455 \kms, and Spur 3 from 450--465 \kms).  Spur 1 has the most
distinctive morphology and velocity structure.  The tip of Spur 1,
seen best in the channel maps from 415--430 \kms, is 350 pc away from
the spiral arm and appears to be a coherent and dense region of radius
90 pc$^2$.  Integrating over this area and velocity range gives a
total CO flux of 1.69 Jy \kms, or a total gas mass of $\sim 1.8 \times
10^6 M_\odot$, and a gas surface density of 71 $M_\odot$ pc$^{-2}$.

Spur 2a and Spur 3 have similar extensions in the channel maps as Spur
1, although their tips are not as distant or distinctive from the base
of each spurs.  We detect CO emission over a length of 475, 420 and
155 pc along Spurs 1, 2a and 3 respectively.  Integrating over these
region we measure total spur masses of 3.4, 5.3 and $\sim$3.9 $\times
10^6 M_\odot$, respectively.  The total area over which there is CO
emission in the spurs ranges from 5.4--9.5$\times$$10^4$ pc$^2$.
Therefore the gas surface densities are 49, 56, $>$72 $M_\odot$
pc$^{-2}$ for spurs 1, 2a, and 3 respectively.  The molecular gas
properties of the spurs are summarized in Table \ref{spurgasprop}.  We
do not detect CO emission from the spur labeled 2b.  Since the spurs
are identified in extinction and the geometry of the dust and stars is
not known, the column density in this spur is likely to be low.  It is
plausible that the spiral arm peak 2 and the slight extension to the
northeast may be the base of spur 2b. Given the lack of CO detection,
we put an upper limit of 6.0$\times$$10^5 M_\sun$ and 6$M_\sun $
pc$^{-2}$ on this spur.

\begin{deluxetable}{cccccc}
\tabletypesize{\scriptsize} 
\tablecaption{Spur Properties: Molecular Gas \label{spurgasprop}} 
\tablewidth{0pt} \tablehead{ \colhead{Spur} & \colhead{$-\Delta\alpha^a$} & \colhead{$\Delta\delta^a$} & \colhead{CO Flux} & \colhead{Mass} & \colhead{$\Sigma$} \\ 
\colhead{} & \colhead{s} & \colhead{$''$} & \colhead{Jy kms$^{-1}$} & \colhead{$10^6 M_\sun$} & \colhead{$M_\sun $ pc$^{-2}$}}
\startdata 
1$^b$& 1.32       & 28.16       & 1.69 & 1.8         & 71 \\   
1    & 1.55-1.32  & 24.66-28.16 & 3.12 & 3.4         & 49 \\ 
2a   & 2.53-2.14  & 16.24-20.69 & 4.86 & 5.3         & 56 \\ 
3    & $\sim$2.89 & $\sim$8.75  & 3.53 & $\simeq$3.9 & $\simeq$72 \\
\enddata
\tablenotetext{a}{The first value is that of the spur base while the second
value is the spur tip.}
\tablenotetext{b}{Values based wholly on the properties of the tip of the spur.}
\end{deluxetable}

\section{Discussion}\label{discussion}

The detection of CO emission in the spurs shows that the spurs are
massive and dense structures with significant amounts of molecular gas
in M51.  The most massive fragment in a spur in our data is at least
as large as some of the largest giant molecular clouds / associations
in the Milky Way \citep{scoville75,sanders85,dame86,sheth08}.  The
mean density in the arm, defined by the arm segment between Spur 1 and
2a extending to the edge of the emission to the southeast, is
300~$M_\sun$ pc$^{-2}$.  The peak density in the arm is a factor of
two or more than this density implying that, overall, the spurs are
significantly weaker relative to the strong spiral arm emission.
Since the spiral arm emission is so bright in M51, the spurs in M51
are likely to be brighter and more massive relative to spurs in other
galaxies.  On-going larger surveys with CARMA may shed light on the
overall properties of spurs (e.g., \citealt{koda08,lavigne08}).
Future higher resolution observations will help delineate the spur
structures and provide detailed kinematics which are needed for
distinguishing between the current models of spur formation and
evolution.

\begin{figure*}
\includegraphics*[width=15.0 cm, angle=-90]{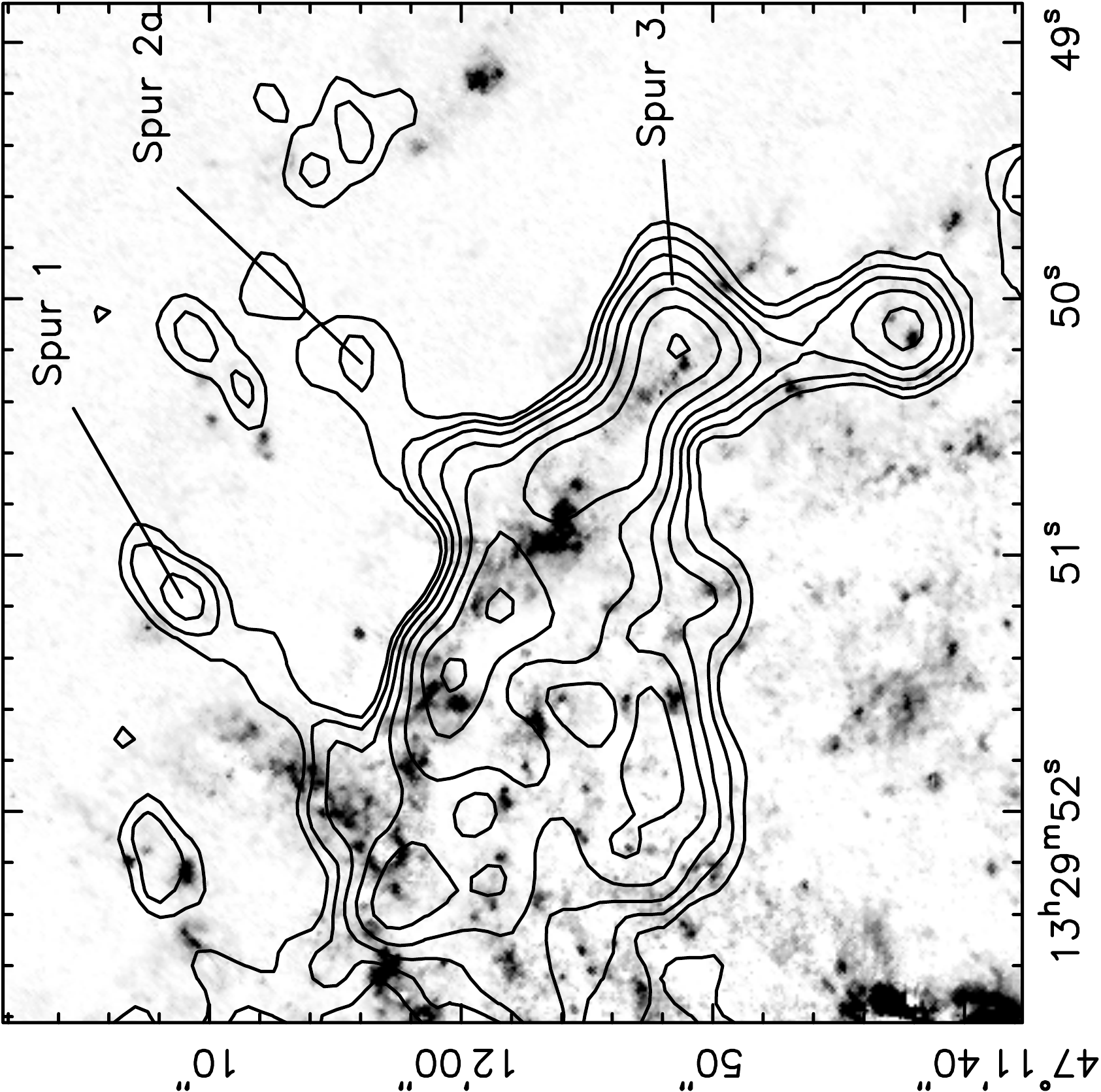}
\caption{\Ha HST image of \citet{scoville01} with CO emission contour
map from OVRO overlaid.  While slightly offset, the H$\alpha$ emission
northeast of spur 1 is morphologically identical to the CO emission
and is thought to be associated with the spur.
\label{ovroHa}}
\end{figure*}

The detection of molecular gas in spurs is significant for shedding
light on the long standing problem of the inter-arm HII regions.  The
widely accepted picture of star formation in spiral arms has been that
clouds converge in the spiral arms, collide and form giant molecular
associations in which star formation is triggered.  Both the newly
formed stars and clouds then travel outwards into the inter-arm
region.  There is an observed offset between the HII regions near the
spiral arm and the peak of the molecular gas density in the arm.  This
same offset has been observed in other galaxies.  \citet{vogel88}
offered an explanation for the offset in terms of a time delay between
the agglomeration of clouds in the spiral arm and the star formation
activity, which is downstream from the spiral arms.  However this
delay becomes increasingly unpalatable for inter-arm HII regions which
are offset more than a few hundred pc from the spiral arm.  The
detection of dense molecular gas in spurs may be the solution since
the spurs extend hundreds of parsecs into the inter-arm area.  When we
examine the structure at the tip of spur 1 we find that it is
kinematically distinct from the base and extended body of the spur.
The total mass in this structure is 1.8$\times$$10^6$\Msun. The
structure itself has a diameter of $\sim$180 pc and a linewidth of
10.4 \kms.  Its virial mass, about 4$\times$$10^6$\Msun, is greater
than the mass measured from the CO flux.  This suggests that the
fragment may not be bound.  If bound, it suggests that the tip has
fragmented from the body of the spur and may be travelling
independently in the inter-arm region.  If unbound, the apparent spur
tip may just be a congregation of lower mass clouds which may
eventually diverge then becoming undetectable at our image depth.
Regardless of the nature of this spur tip, there is star formation
associated with the tip as seen by the presence of H$\alpha$ emission
(see Figure~\ref{ovroHa}) implying that it can be the source of
inter-arm HII regions.  This explanation is also proposed in the
modeling studies by \citet{kim02, kim06}, in which they form clouds of
masses $\sim$7$\times$10$^{6}$ \Msun in the spurs, similar to what we
observe in M51.

\section{Conclusions}\label{conclu}

We present deep, high-resolution CO(1-0) observations of a single field in
M51.  We determine that the extinction features seen in optical images
of M51 \citep{scoville01} are associated with gaseous structures of
mass of a few $10^6$\Msun.  The length of these features is found to
be a few hundred parsecs, consistent with previous measures of these
features in extinction \citep{lavigne06,elmegreen80}.  Given this
separation, the association with H$\alpha$ emission, and a decoupling
of the spur tip from the spiral arm pattern, we find that star
formation in these spurs can result in previously unexplained inter-arm
HII regions.  The origin of these spurs and their importance to the
overall population of inter-arm HII regions requires a wide-field
survey of M51 and possibly other galaxies.

\acknowledgments SC is supported by an NSF Graduate Research
Fellowship.  The Owens Valley Radio Observatory was funded in part by
NSF grant AST 99-81546.

\bibliographystyle{apj} \bibliography{resorbit}

\end{document}